\documentclass[twocolumn,separability,Amsterdam, nobibnotes, aps, pra,showpacs,preprintnumbers]{revtex4-1}
\usepackage{amssymb}
\usepackage{array}
\usepackage{graphicx}

\def\beq{\begin{equation}}
\def\eeq{\end{equation}}
\def\bea{\begin{eqnarray}}
\def\eea{\end{eqnarray}}

\usepackage{color}

\begin{document}

\title{  Holographic Complexity in Gauge/String Superconductors}%

\author{Davood Momeni,$^{1}$ Seyed Ali Hosseini Mansoori,$^{2,3}$ and Ratbay Myrzakulov$^1$ }
\affiliation{$^1$ Eurasian International Center for Theoretical Physics and Department of General Theoretical Physics, Eurasian National University, Astana 010008, Kazakhstan\\
$^2$ Department of Physics, Boston University, 590 Commonwealth Ave., Boston, MA 02215, USA\\
$^3$Department of Physics, Isfahan University of Technology, Isfahan 84156-83111, Iran
}
\email{momeni\_d@enu.kz ; \,\,\ \\
shossein@bu.edu; \,\,\ \\
rmyrzakulov@gmail.com}

\begin{abstract}
Following a methodology similar to \cite{Alishahiha:2015rta}, we derive a
holographic complexity for two dimensional holographic superconductors (gauge/string superconductors) with backreactions. Applying a perturbation method proposed by Kanno in Ref. \cite{kanno}, we study behaviors of the complexity for a dual quantum system near critical points. We show that when a system moves from the normal phase ($T>T_c$) to the superconductor phase  ($T<T_c$), the holographic complexity will be divergent.
\end{abstract}
\pacs{11.25.Tq, 04.70.Bw, 74.20.-z}

\maketitle

\section{Introduction}\label{a}

The anti-de Sitter/conformal field theories (AdS/CFT) correspondence provides us a holographic dual description of the strongly interacting in various fields of physics, especially in condensed matter physics. More precisely, this correspondence establishes a dual relationship between the $d$ dimensional strongly interacting theories on the boundary and the $d + 1$ dimensional weakly coupled gravity theories in the bulk \cite{Maldacena:1997re}. One of the most widely investigated objects is the holographic superconductors. The simple holographic superconductor model dual to gravity theories was made by applying a scalar field and a Maxwell field coupled in an AdS black hole background \cite{HartnollRev,HerzogRev,HorowitzRev}. Then a lot of works have been carried out for investigating holographic superconductors in other complicated gravity theories such as Einstein-Gauss-Bonnet gravity, Horava-Lifshitz gravity, non-linear electrodynamics gravity and so on \cite{r10,r11,r12}.

According to AdS/CFT duality, instability of the bulk black hole leads to exist a conductor and superconductor phase transition in holographic superconductor models. A holographic superconductor/insulator phase transition model was also built at zero temperature \cite{r1}. Then, in Ref. \cite{r2} authors investigated a complete phase diagram for a holographic s-wave superconductor/conductor/insulator system by mixing the conductor/superconductor phase transition with the insulator/superconductor phase transition. 

On the other hand, the entanglement entropy plays a main role in distinguishing different phases and corresponding phase transitions. It is also considered as a useful tool for keeping track of the degrees of freedom of strongly coupled systems. 

In framework of AdS/CFT duality, a holographic method for evaluating the entanglement of quantum systems has been proposed by Ryu and Takayanagi \cite{Ryu:2006bv}. Following this conjecture, the entanglement entropy of CFT’s states living on the boundary of an AdS spacetime is associated with the area of a minimal surface defined in the bulk of that spacetime. Namely, the holographic entanglement entropy of subsystem $A$ with its complement is given by:
\begin{equation} \label{HEE}
S_{A}=\frac{Area(\gamma _{A})}{4G_{d+1}}
\end{equation}
where $G$ and $\gamma_{A}$ are the gravitational constant in the bulk and the $(d-1)$-minimal surface extended into the bulk with the same boundary $\partial A$ of subsystem $A$, respectively. Recently, the behavior of entanglement entropy for holographic superconductor models have been studied in investigating conductor/superconductor phase transitions \cite{r3}-\cite{Erdmenger:2015spo}.  We also obtained an exact form of the holographic entropy due to an 
advantage of approximate solutions inside normal and superconducting phases with backreactions for 2D holographic superconductors \cite{Momeni:2015iea}. 

Recently, Susskind has found a new quantity releted to {\it complexity}  in CFTs which is dual to a volume of a codimension one time slice in anti-de Sitter (AdS) \cite{Susskind:2014rva1,Susskind:2014rva2}. The time slice connect two boundaries dual to the thermofield doubled CFTs, through the Einstein-Rosen bridge \cite{Stanford:2014jda}. Following Refs. \cite{Susskind:2014rva1,Susskind:2014rva2}, the holographic complexity can be defined as:
\begin{equation}\label{HC}
\mathcal{C}_A= \frac{V(\gamma)}{8\pi R G_{d+1}},
\end{equation}
where $R$ and $V(\gamma)$ are the radius of the curvature and the volume of the part in the bulk geometry enclosed by the minimal hypersurface appearing in the computation of entanglement entropy, respectively.
This quantity can be carried out for certain holographic models \cite{Alishahiha:2015rta}. Today it is used to get more information about the quantum systems in black hole physics and cosmology \cite{Barbon:2015soa,Barbon:2015ria}. In the framework of AdS/CFT correspondence,  holographic complexity in bulk might provide a dual description of the fidelity defined in quantum information \cite{Alishahiha:2015rta,r5}. Beacuse the fidelity is purely a quantum information concept, it would be a great advantage if one use it to characterize the quantum phase transitions \cite{r6,r7}.  Moreover, if one extend the fidelity to thermal states, the leading term (fidelity susceptibility) of the fidelity between two neighboring thermal states is simply the specific heat \cite{r8}. Roughly speaking, the fidelity approach is a powerful tool to talk about quantum (or thermal) phase transitions \cite{r6,r7}.
 In this paper, we will investigate behaviors of phase transitions for 2D holographic superconductors by calculating the holographic complexity. The case of 2D holographic superconductivity of our interest is based on the $AdS_{3}/CFT_{2}$ correspondence.  We also consider the influence of the backreaction on the dynamics of perturbation in the background spacetime. Then by applying the domain wall approximation analysis \cite{Albash2012}, we will get the holographic complexity which would be divergent at critical points.

The paper is organized as follows. In the next Section, we will have a brief glance at 2D holographic superconductors. In Sections \ref{S2} , we will calculate the holographic complexity to analyze the phase transition in such superconductors.
 Finally, Section \ref{S5} is devoted to conclusions.
\section{2D holographic superconductor with backreactions}\label{S1}
 In this section, we begin with a brief review of 2D holographic superconductor away from the probe limit by considering the backreaction. The dual gravity description of these superconductors is defined by the following action
\cite{ren}:
\begin{eqnarray}\label{action}
&&S=\int d^3
x\sqrt{-g}\Big[\frac{1}{2\kappa^2}(R+\frac{2}{l^2})-\frac{1}{4}F^{ab}F_{ab}\\&&\nonumber-|(\nabla-i
e A)\psi |^2-m^2|\psi|^2\Big].
\end{eqnarray}
in which $\kappa^2=8 \pi G_3$, $F_{\mu\nu}=\nabla_{\mu}A_{\nu}-\nabla_{\nu}A_{\mu}$ where $A_{\mu}$ are Maxwell's fields, and $e$ and $m$ represent the charge and the mass of the scalar field $\psi$.  In order to regard the effect of the backreaction of the holographic superconductor, we consider a metric ansatz as follows:
\begin{eqnarray}\label{ansatz1}
ds^2=-f(r)e^{-\beta(r)}dt^2+\frac{dr^2}{f(r)}+\frac{r^2}{l^2}dx^2~.
\end{eqnarray}
and the electromagnetic field and the scalar field can be chosen as: 
\begin{eqnarray}\label{ansatz2}
A_{\mu}dx^{\mu}=\phi(r)dt,\ \ \psi\equiv\psi(r).
\end{eqnarray}
Furthermore we consider $\psi$ as a real function without loss of generality. The Hawking temperature of this black hole, which is equivalent to the temperature of the CFT, is given by:
\begin{eqnarray}\label{temperature}
T=\left.\frac{f'(r)e^{-\beta(r)/2}}{4
\pi}\right|_{r=r_+}~.
\end{eqnarray}
Employing the ansatz (\ref{ansatz1}) and (\ref{ansatz2}), the equations of motion can be easily obtained by the following relations \cite{plb}:
\begin{widetext}
\begin{eqnarray}\label{eom}
&&\psi ''(r)+\psi '(r)
\left[\frac{1}{r}+\frac{f'(r)}{f(r)}-\frac{\beta
'(r)}{2}\right]+\psi (r) \left[\frac{e^2 \phi (r)^2
e^{\beta
(r}}{f(r)^2}-\frac{m^2}{f(r)}\right]=0~,\nonumber\\&&
\phi''(r)+\phi '(r) \left[\frac{1}{r}+\frac{\beta '(r)}{2}\right]-\frac{2 e^2 \phi(r) \psi( r)^2}{f(r)}=0~,\nonumber\\&&
f'(r)+2 \kappa ^2r \left[\frac{e^2 \phi (r)^2 \psi
(r)^2 e^{\beta (r)}}{f(r)}+f(r) \psi
'(r)^2+m^2 \psi (r)^2+\frac{1}{2}
 e^{\beta (r)} \phi '(r)^2\right]-\frac{2r}{l^2}=0,\nonumber\\&&
\beta '(r)+ 4 \kappa ^2 r \left[\frac{e^2 \phi(r)^2
\psi(r)^2 e^{\beta (r)}}{f(r)^2}+\psi
'(r)^2\right]=0.
\end{eqnarray}
\end{widetext}
Here the prime denotes the derivative with respect to $r$. In Ref. \cite{plb}, to find the effect of the backreaction on the scalar condensation, Yunqi Liu {\it et al}  did numerical calculations. They showed that numerical results in solving equation (\ref{eom}) for various values of the backreaction $\kappa^{2}$ leads to drop the critical temperatures ($T_c\sim \rho$, where $\rho$ is the dual chemical potential in CFT) consistently  when the backreaction grows. It means that the backreaction makes the condensation harder to occur. They also confirmed that the gap of the condensation operator $<\mathcal{O}_{+} >$ becomes bigger if the backreaction increases. 
Moreover near the phase transition the condensation operator shows itself a behavior like  $<\mathcal{O}_{+} >\sim \sqrt{1-\frac{T}{T_c}}$ which is the same as expected results from mean field theory. The exponent $\frac{1}{2}$ implies that the second order phase transition occurs. Furthermore, at the zero temperature limit $T=0$, the condensate  $<\mathcal{O}_{+} >$ goes to infinity that it justifies the same results from the BCS theory.
(see  Ref. \cite{Momeni:2013waa} for an analytic description of phase transitions).
\section{ Holographic complexity and phase transitions}\label{S2}
Now, we study the holographic complexity for 2D holographic superconductors. By defining $z=r_{+}/{r}$ in the Poincare’s coordinate, and replacing it into the metric (Eq. \ref{ansatz1}), the volume function yields: 
\begin{eqnarray}
&&V(\gamma)=\frac{r_{+}^2}{l}\int \frac{x(z) dz}{z^3\sqrt{f(z)}}\label{V}.
\end{eqnarray}
in order to find $x(z)$ in above formula,
 let us consider an entangling region (subsystem A) in the shape of a strip \cite{Ryu:2006bv, syed}. The minimal surface $\gamma_{A}$ is a one dimensional hypersurface (geodesic) at $t = 0$ when  Eq. (\ref{HEE}) is employed. It should be noted  that none of the coordinates $(z; x)$ is independent of the other. Therefore, considering $z$ as a function of $x$, the surface area becomes:
\begin{eqnarray}
&&A(z(x))=r_{+}\int \frac{dx}{z^2}\sqrt{\frac{z'^2}{f(z)}+\frac{z^2}{l^2}}\label{A}.
\end{eqnarray}
It is noteworthy that we have not imposed the minimality condition on the surface area yet. We now use the Hamiltonian approach to minimize the surface area. Therefore, we get to
the first order differential equation as follows,
\begin{eqnarray}
&& \frac{z'^2}{f(z)}+\frac{z^2}{l^2}=(Cl^2)^2.
\end{eqnarray}
 Defining a turning point $z_{*}$ such that $z'|_{z=z_{*}}=0$, one can obtain the following minimal path for $x(z)$ as:
\begin{eqnarray}
&& x(z)=l\int\frac{dz}{\sqrt{f(z)}\sqrt{z_{*}^2-z^2}}\label{x(z)}.
\end{eqnarray}
Let us regard $f(z)=f_0(z)+\epsilon^2 f_2(z)+...$,  $\beta(z)=\epsilon^2\beta_2(z)+...$, and  $\mu=\mu_0+\epsilon^2\delta\mu_2+...$ where $\epsilon$ is interpreted as the coherence length in a superconductor which is considered small  and $\delta\mu_2\ll\mu_0$ \cite{kanno,Herzog-2010,Ge:2011cw}. It is clear that when $\mu\to\mu_0$, the order parameter $\epsilon^2\to0$. The phase transition occurs  at the critical value $\mu_c=\mu_0$. The equation of motion for $f(z)$ is solved at zeroth order by 
\begin{equation}
f_0(z)=\frac{r_{+}^2}{l^2 }(z^{-2}-1)+\kappa^2\mu^2\log z
\end{equation} 
Near the critical point, $T\sim T_c$, we have  $\psi\sim\psi_1(z)\approx \epsilon z^{2}$. Therefore, one can take the following expression for $f_2(z)$ when $m^2=0$.
\begin{eqnarray}\label{f2}
&&f_2(z)=-\kappa^2 \left( 2C_2+C_1\mu \right) z
\end{eqnarray}
where $C_1$ and $C_{2}$ are integral constants. 
 Following above considerations, the Eq. (\ref{x(z)}) can be divided into two parts.
\begin{equation}
x=l\int\frac{dz}{\sqrt{z_{*}^2-z^2}\sqrt{f_0(z)}}-\frac{\epsilon^2 l}{2}\int\frac{f_2(z)dz}{\sqrt{z_{*}^2-z^2}f_0(z)^{3/2}}
\end{equation}
We assume that effects of backreactions to be small, so one could expand the above equation in powers of $\kappa^2$ as follows:
\begin{eqnarray}\label{xz}
&&x(z)=\frac{l^2}{r_{+}}\sum_{n=0}^{\infty}c_n^{1/2}J_{n}^{1/2}(z_{*}|z)\\&&\nonumber+\frac{\epsilon^2 \kappa^2 l^4(2C_2+C_1\mu)}{2r_{+}^3}
\sum_{n=0}^{\infty}c_n^{3/2}J_{n}^{3/2}(z_{*}|z)
\end{eqnarray}
where $c_n^m=\frac{(m)_n}{n!}\Big(\frac{\kappa l \mu}{r_{+}}\Big)^{2n}$, $ (a)_n = \frac{(a+n-1)! }{(a-1)!}$ is the Pochhammer symbol \cite{M. Abramowitz} and the auxiliary integrals are defined as:
\begin{eqnarray}
&&\nonumber J_{n}^{1/2}(z_{*}|z)\equiv \sum_{k=0}^{\infty}\sum_{p=0}^{\infty}\frac{(n+1/2)_k(1/2)_p}
{k!p!z_{*}^{p+1}}L(2(n+k)+1+p||n)\\
&&\nonumber J_{n}^{3/2}(z_{*}|z)\equiv \sum_{k=0}^{\infty}\sum_{p=0}^{\infty}\frac{(n+3/2)_k (1/2)_p}{k!p!z_{*}^{p+1}}L(2(n+k)+4+p||n).
\end{eqnarray}
where 
\begin{equation}
L(M||N)\equiv \int z^{M}(\log z)^N dz
\end{equation}
is the Log-integral which satisfies the following recursion relation.
\begin{equation}
(M+1) L(M||N)+NL(M||N-1)=z^{M+1}(\log z)^N
\end{equation}
On the other hand, we can employ the "Domain Wall" method in calculating holographic complexity.
Domain wall approach was
proposed to investigate some aspects of the holographic entanglement entropy along renormalization
group (RG) trajectories \cite{Albash2012}.
Thus the expression for holographic complexity (\ref{V}) can be written in the below form:
\begin{equation}
V(\gamma)=\frac{r_{+}^2}{l}\Big(\int_{z_*}^{z_{DW}}\frac{x(z) dz}{z^3\sqrt{f(z)}}+\int_{z_{DW}}^{z_{UV}} \frac{x(z) dz}{z^3\sqrt{f(z)}}\Big)\label{VDW}.
\end{equation}
By substituting (\ref{xz}) into (\ref{VDW}), we arrive at:
\begin{widetext}
\begin{eqnarray}
&&V(\gamma)=\frac{r_{+}^2}{l^2}\Big[\int_{z_*}^{z_{DW}}\frac{dz}{z^3}\Big(\sum_{n=0}^{\infty}d_n^{1/2}J_n^{1/2}(z|z_{*})+\sum_{n=0}^{\infty}d_n^{3/2}J_n^{3/2}(z|z_{*})\Big)\\&&\nonumber\times
\Big(\sum_{m=0}^{\infty}d_m^{1/2}\frac{(\log z)^m}{(z^{-2}-1)^{m+1/2}}+\sum_{m=0}^{\infty}d_m^{3/2}\frac{(\log z)^m}{(z^{-2}-1)^{m+3/2}}
\Big)\\&&\nonumber+\int_{z_{DW}}^{z_{UV}}\frac{dz}{z^3}\Big(\sum_{n=0}^{\infty}d_n^{1/2}J_n^{1/2}(z|z_{*})+\sum_{n=0}^{\infty}d_n^{3/2}J_n^{3/2}(z|z_{*})\Big)\\&&\nonumber\times
\Big(\sum_{m=0}^{\infty}d_m^{1/2}\frac{(\log z)^m}{(z^{-2}-1)^{m+1/2}}+\sum_{m=0}^{\infty}d_m^{3/2}\frac{(\log z)^m}{(z^{-2}-1)^{m+3/2}}\Big)\label{VDW2},
\end{eqnarray}
\end{widetext}
in which $d_n^{1/2}=\frac{l^2}{r_{+}}c_n^{1/2},d_n^{3/2}=\frac{\epsilon^2 \kappa^2 l^4(2C_2+C_1\mu)}{2r_{+}^3}c_n^{3/2}$. Considering the following integral function,
\begin{eqnarray}
&&K_{nm}^{\alpha\gamma}(z_{<},z_{>})=\int_{z_{<}}^{z_{>}}\frac{J_{n}^{\alpha}(z|z_{*})(\log z)^m}{z^3(z^{-2}-1)^{n+\gamma}}dz
\end{eqnarray}
with $(\alpha,\gamma)=(1/2,3/2)$ and
$(z_{<},z_{>})=(z_{*},z_{DW},z_{UV})$, one can rewrite down the Eq. (\ref{VDW2}) as the following explicit form.
\begin{widetext}
\begin{equation}\label{EQ1}
V(\gamma)=\frac{r_{+}^2}{l^2}\sum_{n=0}^{\infty}\sum_{m=0}^{\infty}\sum_{\alpha,\beta=1/2,3/2}D_{n}^{\alpha}D_{m}^{\beta}\Big(K_{nm}^{\alpha\beta}(z_{*},z_{DW})+K_{nm}^{\alpha\beta}(z_{DW},z_{UV})
\Big)\label{V}
\end{equation}
\end{widetext}
where coefficients can be defined as:
\begin{equation}
D_{n}^{\alpha}=\frac{g_n^{1/2}}{r_{+}^{2(n+1/2)}}\delta_{\alpha,1/2}+\frac{\kappa^2\epsilon^2(2C_2+C_1\mu) h_{n}^{3/2}}{r_{+}^{2(n+3/2)}}\delta_{\alpha,3/2}.
\end{equation}
It should be noted that the thermal part involved in the temperature and dual chemical potential are kept out the coefficients $g$ and $h$.  We can express the holographic complexity
 in terms of the temperature $T$, critical parameters like $T_c$, $\mu_c$ and dual quantity as $\mu$. Note that from the coherence length near $T_{c}$ behaves as:
 \begin{equation}
  \epsilon\approx<\mathcal{O}_{+} >\approx\sqrt{\mu-\mu_c}\approx\sqrt{T_c-T}
 \end{equation}
Furthermore we mention here that CFT temperature needs to be modified by including the backreaction of fields on the BTZ black hole background. According to Eq. (\ref{temperature}), the first order correction to the temperature is given by the following expression:
\begin{equation}\label{TT1}
T\approx T_0\Big[1-\frac{ {\epsilon}
^{2}}{4}\,{\frac { \left( 2\,\pi \,\beta_{{2}} \left( z
 \right) +r_{+}f'_{{2}} \left( z \right)T_0^{-1}  \right)}{\pi }}|_{z=1}\Big]
\end{equation}
where $T_0=\frac{f_0'(r_{+})}{4\pi}$ is Hawking temperature for pure BTZ black hole.  When $\mu_0=\mu_c$, the above Hawking temperature tends to the critical point $T_c$ where the order $\epsilon^2\to0$. By putting the $f_2(z)$ (Eq. (\ref{f2})) and 
\begin{equation}
\beta_2(z)=-16\kappa^2\int z^3 dz \Big(1-\frac{l^4z^2\mu^2r_{+}^2\ln(\frac{z}{r_{+}})^2}{4\Big(\kappa^2\mu^2\ln(\frac{z}{r_{+}})+r_{+}^2(1-z^2)
\Big)^2}
\Big)
\end{equation}
 in Eq. (\ref{TT1}), we have:
 \begin{equation}
T\approx T_0\Big[1+\frac{ {\epsilon}
^{2}}{4\pi}( 2\,\pi \,C_{{1}}+{\frac {r_{+}{\kappa}^{2} \left( 2\,C_{{2}}+C_{{1}}\mu
 \right) }{T_{{0}}}})\Big]
\end{equation}
It is also easy to calculate the horizon $r_{+}$ as a function of other variables as:
 \begin{equation}
r_{+}\approx\frac{2\pi T_0}{\kappa^2(2C_2+C_1\mu)}\Big[\frac{2\delta\mu_2}{\mu-\mu_0}\Big(\frac{T}{T_0}-1
\Big)-C_1
\Big]\label{r}.
\end{equation}
Now we want to show that near the critical point, when the system moves to the superconductor phase, $ \mathcal{C}(\frac{T}{T_0},\mu,\mu_c)$ goes to infinity. In other words, the holographic complexity will be divergent as the fidelity susceptibility near the critical points. For this purpose, let us consider the leading order term of  $\mathcal{C}(\frac{T}{T_0},\mu,\mu_c)$ when one takes $m=n=0$ in Eq. (\ref{EQ1}). Thus we have:
\begin{widetext}
\begin{eqnarray}
&&\mathcal{C}(\frac{T}{T_0},\mu,\mu_c)|_{T\to T_c}\approx\frac{1}{ l^3 \kappa^2}\Big[
p_{00}^{1/2,1/2}(z_{<},z_{>})\Big(\frac{2\pi T_0}{\kappa^2(2C_2+C_1\mu)}\Big[\frac{2\delta\mu_2}{\mu-\mu_c}\Big(\frac{T}{T_0}-1
\Big)-C_1
\Big]\Big)^{2}\\&&\nonumber+\frac{\kappa^2(2C_2+C_1\mu)(\mu-\mu_c)}{\delta\mu_2}\Big(
p_{00}^{1/2,3/2}(z_{<},z_{>})+
p_{00}^{3/2,1/2}(z_{<},z_{>})\Big)\\&&\nonumber
+
\frac{\kappa^4(\mu-\mu_c)^2(2C_2+C_1\mu)^2 }{{\delta\mu_2}^2}p_{00}^{3/2,3/2}(z_{<},z_{>})\Big(\frac{2\pi T_0}{\kappa^2(2C_2+C_1\mu)}\Big[\frac{2\delta\mu_2}{\mu-\mu_c}\Big(\frac{T}{T_0}-1
\Big)-C_1
\Big]\Big)^{2}
\Big]
\end{eqnarray}
\end{widetext}
It is obvious that when $\mu \to \mu_{c}$, the first term goes to infinity and other leading terms have no singularities. In other words,  entering obliquely scalar hair into the system of a normal black hole at $\mu_c$ leads to be the divergent the holographic complexity of hairy BTZ black hole by varying discontinuously in temperature from $T>T_c$ to $T<T_c$.  It is interesting in proving analytic result  by applying numerical methods in future works.
 \section{Conclusion}\label{S5}
Phase transition is a change in a feature of a physical system, often involving the absorption or emission of energy from the system, resulting in a transition of that system from one state to another state. The type-II superconductor is a type of phase transition of strongly coupled system from a normal phase to a superconductor phase, requiring free energy. We need a condensate to make this phase transition. In gauge-gravity picture, we address phase transition by investigating a weakly coupled gravitational theory in bulk. In this work we studied phase transition in a two dimensional holographic superconductor with the holographic view of complexity. For this purpose,  we used the domain wall method to calculate the holographic complexity. Our result shows that this quantity is singular at critical chemical potential. Its means that singularities of the complexity happen at normal/superconductor phase transition points for 2D holographic superconductors. 




\end{document}